\documentclass[A4paper, 11pt]{article}
 
\usepackage{array}
\usepackage{multirow}
\usepackage[fleqn]{amsmath}
\usepackage{graphicx}
\usepackage{tikz}
\usetikzlibrary{snakes}
\usepackage{float}
\usepackage{amssymb}
\usepackage[latin1]{inputenc}
\usepackage{hyperref}
\usepackage{epstopdf}

\newcommand{\minocc}{\textrm{minOcc}}
\newcommand{\maxocc}{\textrm{maxOcc}}
\newcommand{\maxp}{\textrm{max}}
\newcommand{\noise}{\textit{noise}_{c_n}}

\begin{document}

\title{Tire Noise Optimization Problem: a Mixed Integer Linear Program Approach}
\author{Matthias Becker \\ {\small FG Simulation, Welfengarten1, Universit\"{a}t Hannover, Germany} \\
{\small \url{xmb@sim.uni-hannover.de}} \and Nicolas Ginoux \\ 
{\small Universit\'e de Lorraine, CNRS, IECL, F-57000 Metz, France} \\
{\small \url{nicolas.ginoux@univ-lorraine.fr}}\and S\'ebastien Martin and Zsuzsanna R\'oka \\ {\small Universit\'e de Lorraine, LCOMS,  F-57070 Metz, France} \\
{\small \url{{sebastien.martin, zsuzsanna.roka}@univ-lorraine.fr}}}
\date{}
\maketitle

\begin{abstract}
We present a Mixed Integer Linear Program (MILP) approach in order to model the non-linear problem of minimizing the tire noise. We first take more industrial constraints into account than in a former work of the authors. Then, we associate a Branch-and-Cut algorithm to the MILP to obtain exact solutions. We compare our experimental results with those obtained by other methods. 
\end{abstract}

\section{Introduction}\label{sec:intro}
The reduction of tire noise in the vehicle interior has been a major field of research in the tire industry for many years. After the tire engineers have determined every aspect of a new tire prototype concerning the driving and safety characteristics, the last degree of freedom is the final design of the so-called pitch sequence. In~\cite{williams}, T. A. Williams provides a historical overview of the approaches used by different companies, based on some U.S. Patent documents, from 1934 to 1994.  First studies show that the noise optimization does not give good results when the number of pitch types is small (a mono-space sequence is the noisiest) and the tire mold is difficult when this number is large. 
A realistic tire size is a sequence of around 60--70 pitches, hence a complete search is not feasible because of an exponential computational time. 
First methods concerned patented single pitch sequences and a randomization of the sequence design. In the 1990's,  some heuristic optimization algorithms were used in~\cite{hoffmeister1998tread}, based on  genetic algorithms. In~\cite{kim2006prediction}, the road surface is also taken into account when modeling the noise of certain pitch patterns.
In~\cite{nakajima2000application}, the authors explain the sound generation mechanisms of a tire as well as the mathematical noise model.
Further works deal with application of intelligent heuristic optimization algorithms (such as genetic algorithms and neural networks) on problems with one pitch sequence, with multiple pitch sequences, and on tires with spikes (cf. \cite{becker2006genetic}, ~\cite{smc2016} and references within those articles).
%
%
An approach based on artificial immune systems combined with genetic algorithms is presented in~\cite{chen2009adaptive} and~\cite{Li2009}.
In~\cite{kim2012image}, a pitch sequence with five pitch types and 50 pitches is optimized using a genetic algorithm. 
There and opposed to other approaches, the noise of the tire pattern is assessed via an image based algorithm.
During the optimization procedure, the only free parameter is the pitch sequence. The quality of each sequence is assessed via a spectral analysis of the candidate sequences.
The optimization of the pattern of a single pitch via particle swarm (PSO) is studied in~\cite{chiu2015application}. 

The tire noise is approximated by a mathematical function, based on Fourier coefficients computed on a space of functions that is not a vector space and hence cannot be linearized  in a natural way. 
Nevertheless, a pitch sequence has to satisfy some constraints (pitch length ratios, number of pitches, etc.), and an Integer Linear Program (ILP) approach can be considered once a convenient noise approximation is defined. 

In a previous work~\cite{smc2016}, we propose a linear approximation of the tire noise and an ILP solving the tire noise optimization problem. 
In that paper, the noise computed for a given sequence is an approximation of the real noise, but the ratio between them is bounded by $\sqrt{2}$, see \eqref{eq:upplowbdcn}. In particular, even if  the best found solution by the ILP  is not always optimal,  the ratio with the optimal noise must be bounded by $\sqrt{2}$. 

In the present paper, we  deepen our study: we present a graph modeling of the problem and  improve the computing time of our ILP by adding some new constraints. We then apply a Branch-and-Cut algorithm in order to find the optimal solution. We conclude with some experimental results.

\section{The Tire Noise Optimization Problem (TNOP)}\label{problem}

\begin{figure}
    \centering
        \includegraphics[scale = 0.12]{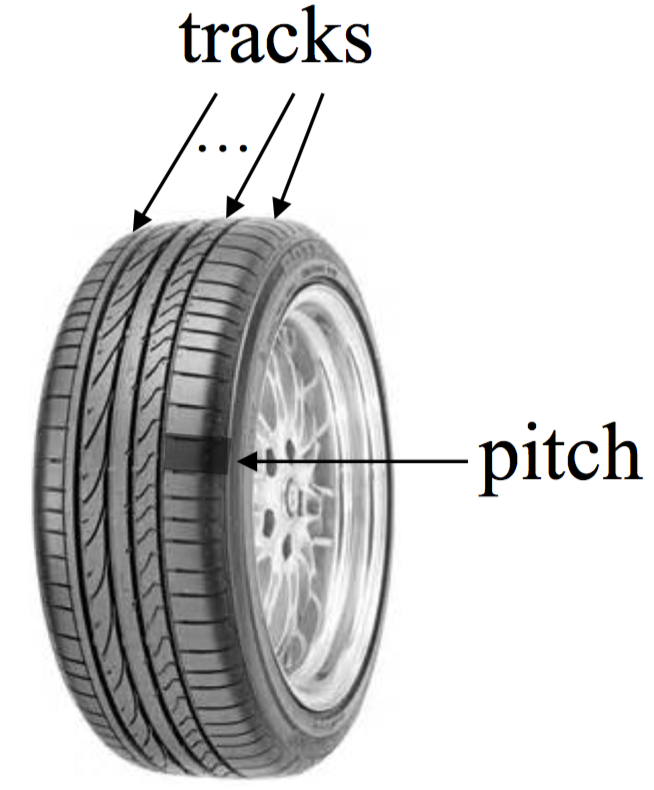}\qquad\qquad
        \includegraphics[scale = 0.80]{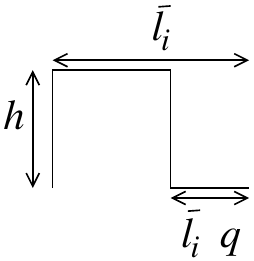}\qquad\qquad
        \includegraphics[scale = 0.75]{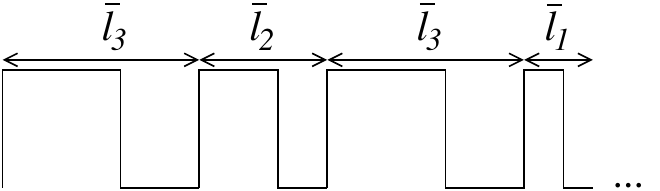}
        \caption {(a) Tracks and pitches, (b) a pitch of size $\bar{l}_i$ and (c) a pitch sequence representing a tire track }
        \label{fig:tire}
    \end{figure}

The surface of a tire consists of different tracks, each track built out of a sequence of pitches of different sizes (see Figure~\ref{fig:tire}a).  A pitch is considered as the juxtaposition of an elevated part and a low part, called groove. The tire noise is  produced by the grooves when coming into contact with the road surface. 
All pitches have the same height $h$ and the groove is a constant fractional part $q$ of the pitch length (Figure~\ref{fig:tire}b).
Let $P:=\{1,\ldots,r\}$ be the finite set of pitch types with lengths $\bar{l}_1<\ldots<\bar{l}_r$. For instance, for $|P| = 3$, the pitch sequence of Figure~\ref{fig:tire}c is modeled by the integer vector $(3, 2, 3, 1, \ldots)$.

In this paper, we only study single track tires, thus the tire profile can be considered as a step function. The tire noise is then roughly defined as a scalar multiple of the largest Fourier coefficient  in absolute value of that function, see Section~\ref{sec:noise} for a formal definition. 

The Tire Noise Optimization Problem (TNOP) consists in finding a pitch sequence producing the lowest noise, respecting given parameters provided by engineers: \medskip

\begin{tabular}{ll}
nbTypes: &  the number of different pitch types \medskip \\
lengthRatios: &  the length ratios of the different pitch types \medskip\\ 
nbPitches: & the total number of pitches composing the tire. \bigskip \\
\end{tabular} 

Notice that the length of each type is not given as an absolute physical number in SI unit, but as a relative length ratio between the types.
Note also that only the total number of pitches is fixed but not the number of pitches of each type, therefore the length of the pitch sequence,  corresponding to the tire circumference,  can vary.  However, this circumference can always be normalized to some standard industrial tire size since the tire noise only depends on the relative pitch lengths. 

To ensure stability in driving and wearing, we also consider  other  constraints such as: \medskip 

\begin{tabular}{lp{12cm}}
ctMinMaxOcc: & the minimal and maximal number of occurrences of each pitch type in a pitch sequence is bounded \medskip\\
ctIncompatibility: & pitches of given types cannot be neighbors in a sequence (\textit{e.g.}\! the shortest and the longest ones)\medskip \\
ctMaxSeq: & the length of a subsequence of pitches of the same type is bounded. \\
\end{tabular} 


\section{Definition of a discrete model}
In this section, we describe a mathematical model for the tire noise and its discretization. In particular, we define length units to transform the relative ratios into integral pitch length thus allowing for an ILP model. 
\subsection{The noise function} \label{sec:noise}The model we study is based on the computation of the Fourier coefficients $a_n, b_n, c_n$ of a given function depending on the profile of the tire.
Recall that, if $f$ is a $T$-periodic function of one real variable, then formally, for all $n\in\mathbb{N}$,
\begin{eqnarray*}
a_n(f)&:=&\frac{2}{T}\int_0^Tf(x)\cos(\frac{2\pi n}{T}x)\,dx\\
b_n(f)&:=&\frac{2}{T}\int_0^Tf(x)\sin(\frac{2\pi n}{T}x)\,dx\\
c_n(f)&:=&\frac{1}{T}\int_0^Tf(x)e^{\frac{-2i\pi n}{T}x}\,dx=\frac{1}{2}(a_n(f)-ib_n(f)).
\end{eqnarray*}
Originally (see e.g.\! \cite{becker2009tread}), the noise produced by a tire with one track is proportional to $$2\max_{n\geq1}(|c_n(f)|),$$ where $f$ is the profile function of the track, see below.

Variable-rescaling does not change the value of the Fourier coefficients: if $f$ is a $T$-periodic function, then for any real positive number $\delta$, the function $f_\delta(x):=f(\delta x)$ is $\frac{T}{\delta}$-periodic and, for every $n\in\mathbb{N}$, we have $c_n(f_\delta)=c_n(f)$ (in particular $a_n(f_\delta)=a_n(f)$ as well as $b_n(f_\delta)=b_n(f)$).
This property has the following important consequence: instead of fixing the circumference $T$ of the tire and let the pitch-lengths vary according to the number of pitches $N$ present on the  track, we may fix the pitch-lengths once for all and let the total circumference vary in terms of $N$; the computed Fourier coefficients (and therefore the noise) will not change under that transformation.

Another important assumption  deals with relative pitch length ratios that we assume to be rational, hence allowing the pitch lengths to be integers (see Section~\ref{sec:units}). 
When we denote a length by an integer, we implicitly assume it is expressed as an integer number of a fixed unit length we do not need to make precise for the computations.

It is worth noticing that several models exist for the profile function. 
For instance, in  \cite{nakajima2000application}, the profile function is described as a finite sum of Dirac functions (actually distributions) at positions $t_1<\ldots<t_N$, where $N$ is the number of pitches and.  In that case the modulus of the Fourier coefficient $c_n$ is given by

$$  |c_n| = \frac{1}{T}\sqrt{ \left(   \sum_{i=1}^N\cos(\frac{2\pi n}{T}t_i)   \right) ^2 + \left(\sum_{i=1}^N \sin(\frac{2\pi n}{T}t_i) \right)^2}, n\in \{1, \ldots, 200\} .$$
 
Remark that, in real life, a tire track can be considered as a sequence of $N=60$ pitches at most. The estimated noise can be computed before the $n=3.5N$th Fourier coefficient.

Trying to calculate the impact of exactly one pitch at a certain position $t_m$ ($1\leq m \leq N$) we can write:

%
 
 \begin{eqnarray*}
C  & =  & \sum_{i=1}^{m-1} \cos(\frac{2\pi n}
 {T}t_i)    + \cos ( \frac{2\pi n}{T}t_m)  +\sum_{i=m+1}^{N} \cos(\frac{2\pi n}{T}t_i) \\
S & = & 
 \sum_{i=1}^{m-1} \sin(\frac{2\pi n}
 {T}t_i)   + \sin(\frac{2\pi n}{T}t_m)  +\sum_{i=m+1}^{N} \sin(\frac{2\pi n}{T}t_i) .
 \end{eqnarray*}

Since $|c_n|=\frac{1}{T}\sqrt{ C^2 + S^2} $,  the  specific harmonic cannot be calculated up to a position $t_m$ because of the squares in the expression, which means we have to multiply e.g.
 the expression for $a_n$ out:
 
$$
 \frac{1}{T^2}\left(   \sum_{i=1}^{m-1}\cos(\frac{2\pi n}{T}t_i)    + \cos(\frac{2\pi n}{T}t_m)  + \sum_{i=m+1}^{N}\cos(\frac{2\pi n}{T}t_i)\right)^2
  $$
which makes everything dependent with everything (the part before position $t_m$ has to be multiplied with the part after  $t_m$, with contribution at $t_m$ itself).

As mentioned in Section~\ref{sec:intro},  several further models have been described in the literature. 
Here we are interested in the model that was proposed in \cite{becker2009tread} and  \cite{smc2016}.
It consists in considering the original profile of the tire as the profile function.
In that case, the Fourier coefficients of a given track may be explicitly computed in terms of the pitch sequence of the track.
From now on, we assume there are $N$ pitches on the given track.
We denote by $l(1),\ldots,l(N)$ the lengths of the successive pitches of a given pitch sequence. 
Then, for any $n\geq1$,
\[a_n(f)=\frac{h}{n\pi}\cdot\sum_{j=1}^N\sin(M_{j,n})-\sin(N_{j,n})\]
\[b_n(f)=-\frac{h}{n\pi}\cdot\sum_{j=1}^N\cos(M_{j,n})-\cos(N_{j,n}),\]
where 
\[M_{j,n}:=\frac{2n\pi }{T}\cdot\left(\sum_{p=1}^{j-1} l(p)+(1-q)l(j)\right)  \textrm{ and }\]  \[N_{j,n}:=\frac{2n\pi }{T}\cdot \sum_{p=1}^{j-1}l(j)=M_{j,n}-(1-q)\frac{2n\pi}{T} l(j).\]
In particular,
\begin{equation}\label{eq:cn}
c_n(f)=\frac{ih}{2n\pi}\cdot\sum_{j=1}^Ne^{-iM_{j,n}}-e^{-iN_{j,n}}.
\end{equation}
It is important to notice a few symmetry properties of the Fourier coefficients.
\begin{itemize}
\item The modulus of each Fourier coefficient $c_n$ is invariant under translation, that is, applying a circular permutation to the pitches  does not change the value of $|c_n|$.
Note however that $|a_n|$ and $|b_n|$ change in general under such transformations.
As a consequence, for the computation of the real noise involving $|c_n|$, we can assume that the first pitch in the sequence is of the first pitch type.
\item Taking the mirror image of a given profile function keeps $a_n$ unchanged while turning each $b_n$ into $-b_n$, in particular it fixes $|a_n|$ and $|b_n|$ -- and thus also $|c_n|$. 

\end{itemize}
In order to handle the noise with an Integer Linear Program (ILP), the authors of~\cite{smc2016} defined the noise of a given one-tracked-tire to be proportional to $$\max_{n\geq1}\left(\max(|a_n(f)|,|b_n(f)|)\right) ,$$ where neither squares nor square-roots enter.
Though \textit{a priori} different, both formulae are roughly equivalent in the following sense: for any real numbers $a,b$,
\begin{equation}\label{eq:upplowbdcn}\max(|a|,|b|)\leq\sqrt{a^2+b^2}\leq\sqrt{2}\cdot\max(|a|,|b|).\end{equation}
Therefore, the noises computed by both formulae cannot be too far from each other, the ratio between both lying in $[1,\sqrt{2}]$.
Moreover, $\max(|a|,|b|)=\sqrt{a^2+b^2}$ if and only if $a=0$ or $b=0$ and $\sqrt{a^2+b^2}=\sqrt{2}\cdot\max(|a|,|b|)$ if and only if $|a|=|b|$.\medskip

\subsection{Fourier precision} In theory, all Fourier coefficients must be computed in order to determine the noise produced by a one-tracked-tire.
However, it is sufficient to compute the $c_n$ coefficients only for $n\leq 200$. A regular tire composed of $N$ pitches has a peak at the frequency of $N$ as well as at multiples of $N$, so called harmonics. Humans are able to hear only frequencies between 20 Hz and 20 kHz, so only the first harmonics of the tire signal are  in the main human audible frequency range, and at most the second harmonics come close to the 20kHz frequency border. Furthermore, higher frequencies of a regular signal are decreasing quickly, so there is no need to compute more than 1.5 to 3.0 times $N$ coefficients, around 200 in the case of a realistic tire pattern with $N=60$ pitches. 

Moreover, because of the very specific profile functions we deal with, it is an easy consequence of \eqref{eq:cn} that, for every $n\geq1$,
$$|c_n(f)|\leq\frac{Nh}{n\pi},$$
where $h>0$ is the pitch height.
In particular, if $N$ is fixed, then $|c_n|$ (as well as $|a_n|$ and $|b_n|$) is an $O\left(\frac{1}{n}\right)$ and thus is small for large $n$.


\subsection{Units and integer pitch lengths}\label{sec:units}
Recall that the set of different pitch types $P=\{1, \ldots, r\}$ is finite and $\bar{l}_1<\ldots<\bar{l}_r$ are the different pitch lengths.
Recall that the ratio of any two pitch lengths is supposed to be rational and hence we can assume the existence of a largest length (called unit) dividing each pitch length into an integral number of units.
Formally, let $\bar{l}_i=\frac{p_i}{q_i}\cdot \bar{l}_r$ with $p_i,q_i\in\mathbb{N}$ and $\mathrm{gcd}(p_i,q_i)=1$ for all $1\leq i\leq r$ (note that $p_r=q_r=1$). We define the  unit length  by
$u:=\frac{\bar{l}_r}{\mathrm{lcm}\left(q_1,\ldots,q_r\right)}$,
where $\mathrm{lcm}\left(q_1,\ldots,q_r\right)$ is the least common multiple of the integers $q_1,\ldots,q_r$.
From now on, we study tires with reduced pitch lengths $l_1<\ldots<l_r$, where $l_i:=\frac{\bar{l}_i}{u}\in\mathbb{N}$  for all $i\in\{1,\ldots,r\}$.
We call $l_{\min}:=N\cdot l_1\in\mathbb{N}$ and $l_{\max}:=N\cdot l_r\in\mathbb{N}$ the minimal and maximal tire-length,
respectively. 
From the industrial perspective, the  sequence of pitch lengths is given as a sequence of ratios with respect to the smallest pitch length. 
Most of the results presented in Section~\ref{sec:results} concern ratios $\{1.0, 1.25, 1.5\}$ reduced to lengths $\{4, 5, 6\}$ in the ILP (here $u=0.25$).   Notice also that we do not study tires of a fixed length $T$ but tires with lengths varying from $l_{\min}$ to $l_{\max}$.

\section{Optimization models}

The results given in the previous section allow to propose a linear optimization model. In this section, we highlight  the link with scheduling, graph and mathematical programming.

\subsection{Graph presentation}\label{sec:graph}
\begin{figure}
\centering
\includegraphics[scale=0.8]{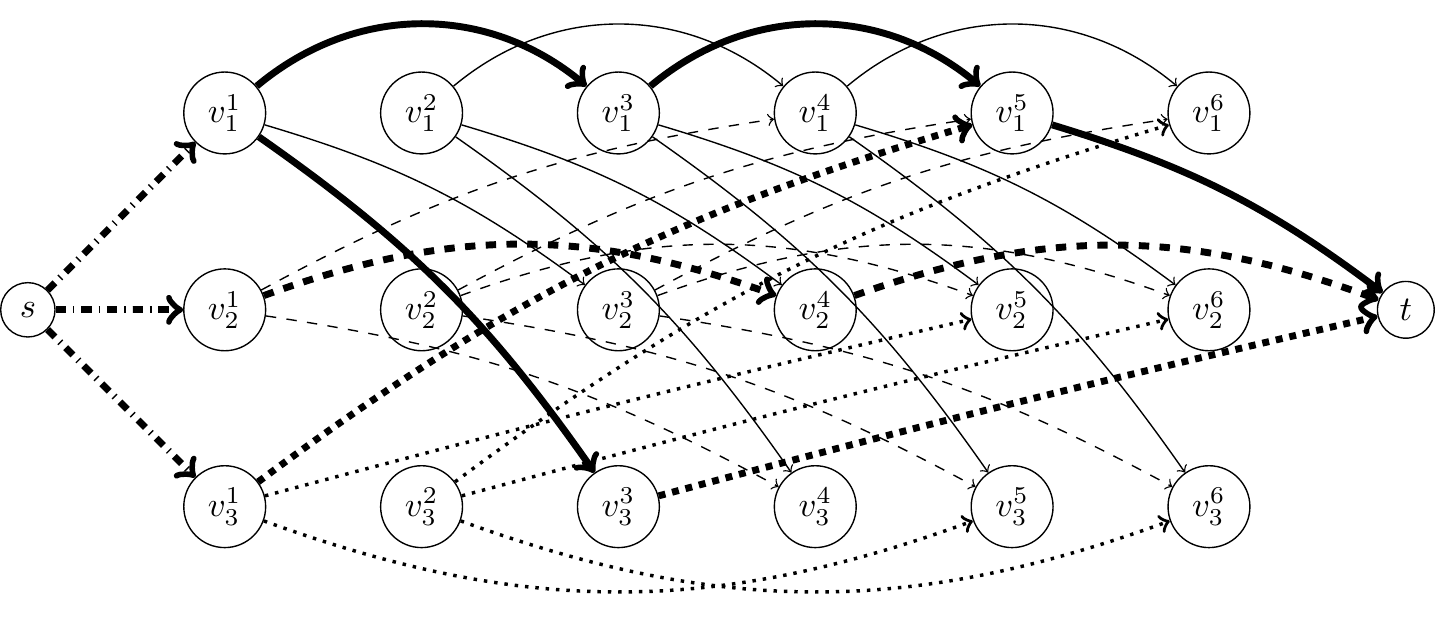}
\caption{Graph model for a tire of length 6.}
\label{fig:graph}
\end{figure}

For the sake of clarity, we consider here tires of length $T=l_{\max}$. From Section~\ref{sec:ilp} on, we let the tire length vary and solve the TNOP for each tire size. Recall that a tire track is a  sequence of consecutive pitches of different sizes assumed to be integers. Then, on each position, a pitch can or cannot start. 
A sequence of pitches can be seen as a scheduling problem on a single machine where the pitches are the jobs and the criteria to minimize depends on the beginning of each pitch. This problem also can be modeled as a graph problem. 
Let $T$ denote the length of a tire expressed as an integral number of the above units and let $G=(V,A)$ be the oriented  graph defined as follows: 
\begin{itemize}

\item $V= \{ v_p^i \mid \forall i\in \{1, \ldots, T\} ; \forall p \in \{1, \ldots, r\} \} \cup \{s, t\}$ where $i$ denotes a position on the tire and $p$ denotes a pitch type,

\item $A=A_s \cup A_{\mathrm{int}} \cup A_t$, where 
\begin{eqnarray*}
A_s & = &  \{(s, v_p^1) \mid p\in \{1, \ldots, r\} \}, \\
A_{\mathrm{int}} & = &\{(v_p^i, v_{p'}^{i+\mathrm{length(p)}}) \mid \forall i\in \{1, \ldots, T \}; \forall p, p' \in \{1, \ldots, r\}, \\ 
 &  & \phantom{\{(v_p^i, v_{p'}^{i+\mathrm{length(p)}}) \mid \forall   }i+\mathrm{length(p)} \leq T\}, \\
A_t & = & \{ (v_p^{T-\mathrm{length(p)}}, t) \mid p\in \{1, \ldots, r \}\}.  
\end{eqnarray*}

\end{itemize}
An example for  a tire of length 6 built of pitches of  lengths 2, 3 and 4 is presented on Figure~\ref{fig:graph}. To each arc $\mathsf{a}$ starting at $v_p^i$  in $A$ and each $k\in \mathbb{N}$, we associate a weight $(w_a^k(\mathsf{a}), w_b^k(\mathsf{a}))$, corresponding to the $a_k$- and $b_k$-contributions to the noise produced by a pitch of type $p$ starting at position $i$. Actually, $w_a^k(\mathsf{a}) = A_{k,i,0,p}$  and $w_b^k(\mathsf{a}) = B_{k,i,0,p}$ according to the definitions of Section~\ref{sec:ilp}. 

Let $\mathsf{P}$ be the set of all paths of $N$ pitches starting at $s$ and ending at $t$. 
Then, the TNOP consists in finding a path $\mathsf{p}\in \mathsf{P}$ such that the maximum noise $\max_k\{{|\sum_{\mathsf{a}\in \mathsf{p}}{w_a^k(\mathsf{a})}|, |\sum_{\mathsf{a}\in \mathsf{p}}{w_b^k(\mathsf{a})}|}\}$ is minimal.

\noindent For the example of Figure~\ref{fig:graph}, for
\begin{itemize}

\item  $N=3$ :  $\mathsf{P} = \{(s,v_1^1)(v_1^1,v_1^3)(v_1^3,v_1^5)(v_1^5,t) \}$

\item  $N=2$ :  $\mathsf{P} = \{(s,v_1^1)(v_1^1,v_3^3)(v_3^3, t), \\ \phantom{N=2 :  \mathsf{P} =  ii}(s,v_2^1)(v_2^1, v_2^4)(v_2^4,t), \\ \phantom{N=2 :  \mathsf{P} = ii}(s,v_3^1)(v_3^1,v_1^5)(v_1^5, t) \} $.

\end{itemize}
 
We propose in the next section an integer linear program based on the path problem defined above.\medskip

\subsection{Mixed Integer Linear Program (MILP)}\label{sec:ilp}
We propose a MILP to solve the problem for a fixed tire length between  $l_{\min}$ and $l_{\max}$, considering hence some empty unit slots at the end of the sequence. Recall that the noise we compute differs from the real noise by a factor that is at most $\sqrt{2}$, as described in Section~\ref{sec:noise}. 

We use the following notations:
\begin{itemize}

\item $N$ is the number of pitches,

\item $P=\{1, 2, \ldots, r\}$ is the set of pitch types and $l_1 < l_2 < \ldots < l_r$ are the reduced pitch lengths, supposed to be integers,

\item $L = \{0,...,l_{\max}-l_{\min}\}$ denotes the set of possible  numbers of  empty unit slots at the end, $\vert L\vert$ is then the number of possible tire lengths,

\item $L_{p}^j = \{1,...,l_{\max}-l_p-j\}$ denotes the set of positions at which a pitch of type $p\in P$ can start on a sequence with $j$ empty unit slots at the end,

\item $T_{j} = l_{\max}-j$ denotes the length of a tire with $j$ empty unit slots at the end,

\item $K=\{1, \ldots, 3N\}$ where $3N$ denotes the Fourier coefficient precision,

\item $A_{k,j,i,p}$  denotes the contribution to the Fourier coefficient $a_k$  of the function having only one pitch of type $p$ starting on the $i^{\textrm{th}}$ unit and such that the $j$ last units do not bear any pitch. Formally, they are computed as follows:
\[A_{0,j,i,p}=\frac{1}{T_{j}}\cdot h\cdot(1-q)l_p=\frac{(1-q)\cdot h\cdot l_p}{(l_{\max}-j)}\]
and for any $k\geq1$,
$$A_{k,j,i,p}=\frac{h}{k\pi}\cdot\left(\sin\left(\frac{2k((i-1)+(1-q)l_p)\pi}{(l_{\max}-j)}\right)-\sin\left(\frac{2k(i-1)\pi}{(l_{\max}-j)}\right)\right) ;$$
\item $B_{k,j,i,p}$ denotes the contribution to the Fourier coefficient  $b_k$ under the same conditions as for $A_{k,j,i,p}$,  they are computed as follows:
$$B_{k,j,i,p}=-\frac{h}{k\pi}\cdot\left(\cos\left(\frac{2k((i-1)+(1-q)l_p)\pi}{(l_{\max}-j)}\right)-\cos\left(\frac{2k(i-1)\pi}{(l_{\max}-j)}\right)\right) ;$$
\item $\minocc _p$: the minimum number of pitches of type $p$ in a sequence;
\item $\maxocc _p$: the maximum number of pitches of type $p$ in a sequence;
\item $\max _p$: the maximal length of a subsequence composed of  pitches of a same type $p$.\medskip
\end{itemize}

\noindent Let us now  introduce the binary variables
\begin{itemize}

\item 
$x^{p}_i  \in \{0, 1\}$: 
\[ x^{p}_i  =\left\{
\begin{array}{ll} 
1 &\textrm{if a pitch of type }p\textrm{ starts  at position }i\medskip \\
0 &\textrm{otherwise} \\
\end{array}
\right. \]
for all $p\in P$ and $i\in L_p^0$

\end{itemize}

and the continuous variables
\begin{itemize}

\item $z^j \in \mathbb{R}_+$:  the noise produced by a pitch sequence with $j$ empty units at the end, for all $j\in L$,

\item $za^j_k \in \mathbb{R}$: the value  of the $k$-th Fourier coefficient $a_k$ of a given pitch sequence with $j$ empty units at the end, for all $j\in L$ and $k\in K$,

\item $zb^j_k \in \mathbb{R}$: the value  of the $k$-th Fourier coefficient $b_k$ of a given pitch sequence with $j$ empty units at the end, for all $j\in L$ and $k\in K$.

\end{itemize}

Let us fix  the number of empty units $j\in L$ at the end of the sequence. 
The  MILP below solves the TNOP for tire length $T_j$.  Notice that, in order to solve the TNOP, the MILP has to be solved for all $j \in L$.\medskip

\noindent Mixed Integer Linear Program (${\mathcal P}$):
\begin{alignat}{5}
\nonumber & \min z^j \\
\label{ctn1:1}&  za_k^j\leq z^j & & \qquad\forall k\in K,\\
\label{ctn1:2}&  -za_k^j\leq z^j & & \qquad\forall k\in K,\\ 
\label{ctn1:11}&  zb_k^j\leq z^j & & \qquad\forall k\in K,\\
\label{ctn1:21}&  -zb_k^j\leq z^j & & \qquad\forall k\in K,\\ 
\label{ctn1:31}&  za_k^j= \sum_{p\in P} \sum_{i\in L_p^j}  A_{k,j,i,p}x_i^p  & & \qquad\forall k\in K,\\
\label{ctn1:32}&  zb_k^j= \sum_{p\in P} \sum_{i\in L_p^j}  B_{k,j,i,p}x_i^p  & & \qquad\forall k\in K,\\
\label{ctn1:4}&  \sum_{p\in P} x_i^p \le 1  &&\qquad\forall i\in \{1,\ldots,T_j\}, \\
\label{ctn1:5}&  \sum_{p\in P}\sum_{i\in L_p^j} l_px_i^p = T_j, & &\\
\label{ctn1:6}&  \sum_{\stackrel{p=1}{\tiny i-l_p>0}}^r  x_{i-l_p}^p = \sum_{p=1}^r  x_{i}^p, & & \qquad\forall i\in\{1,\ldots, T_j\},\\
\label{ctn1:9}&  \sum_{p\in P} \sum_{i\in L_p^j}x_i^p = N, & & \\
 & x_i^p \in\{0,1\}&&\qquad\forall p\in P, \; \forall i\in \{1,\ldots, T_j-l_{p}+1\}, \\
 &z^j \geq 0&&
\end{alignat}

The objective function and inequalities \eqref{ctn1:1}--\eqref{ctn1:21} ensure that $z^j$ is a maximum of the absolute value of $za_k^j$ and $zb_k^j$ for all $k \in K$.
Equalities \eqref{ctn1:31} and \eqref{ctn1:32} compute the noise contribution of the sequence found by the ILP. 
The inequality  \eqref{ctn1:4} ensures that at most one pitch can start at each position.
The equality \eqref{ctn1:5} ensures that the non-empty part is completely filled with pitches, that is, we take into account all possible combinations having a fixed tire length. 
The equalities \eqref{ctn1:6} prevent a pitch from starting inside another pitch,  \textit{i.e.} there is no overlapping. Remark that, this constraint corresponds to the existence of paths of length $N$  in the graph presented in Section~\ref{sec:graph}. 
The equality \eqref{ctn1:9} ensures that exactly $N$ pitches compose the tire.  

The above $({\mathcal P})$ is a model for the basic problem. We can however easily add some more constraints such as those mentioned in Section~\ref{problem}:

\begin{alignat}{5}
 \label{ctn1:4.5}& \sum_{i\in L_p^j} x_i^p  \geq \minocc_p && \qquad\forall p \in P, \\
\label{ctn1:4.6}& \sum_{i\in L_p^j} x_i^p  \leq \maxocc_p && \qquad\forall p \in P, \\
\label{ctn1:10}& \sum_{k=0}^{\max_p-1} x_{i+kl_p}^p  \leq \maxp_p && \qquad\forall p \in P,\  \forall i\in \{1,\ldots, T_j-l_p+1\},\\
\label{ctn1:7}&  x_i^1+x_{i+l_1}^r \leq 1, & & \qquad\forall i\in \{1,\ldots, T_j-l_r-l_1+1\},\\
\label{ctn1:8}& x_{i}^r+ x_{i+l_r}^1 \leq 1, & & \qquad\forall i\in \{1,\ldots, T_j-l_r-l_1+1\},
\end{alignat}
where

\begin{itemize}

\item inequalities~\eqref{ctn1:4.5} and~\eqref{ctn1:4.6} correspond to the constraint ctMinMacOcc,

\item inequalities~\eqref{ctn1:10} express ctMaxSeq,

\item inequalities~\eqref{ctn1:7} and~\eqref{ctn1:8} translate ctIncompatibility for the incompatibility of the smallest and the largest pitch types.

\end{itemize}

\subsection{Exact Branch-and-Cut algorithm (MILP with B$\&$C)}\label{sec:bc}
Recall that the MILP $(\mathcal{P})$ does not provide an exact solution. 
In this section, we propose an algorithm based on  $(\mathcal{P})$ allowing to find the optimal solution of the TNOP. 

Let $z_{ub}$ be an upper bound for the approximated noise. We initialize $z_{ub}$  by the heuristic algorithm described in \cite{ becker2009tread}. From the MILP $(\mathcal{P})$ we derive a Branch-and-Cut algorithm to ensure that the solution is optimal for the TNOP. During the execution of the Branch-and-Bound algorithm associated with the solving of $(\mathcal{P})$, for each integral feasible solution $\tilde{x}$ found by the Branch-and-Bound, the real noise is computed and denoted by $\noise(\tilde{x})$.  If $\noise(\tilde{x}) < z_{ub} $ then the solution found is better, so that we improve the upper bound by setting $z_{ub} = \noise(\tilde{x})$. As $\tilde{x}$ is momentarily the best solution found, we save it in  $x^*$ and we add the  inequalities 
\begin{alignat}{5}
\label{ctBC} z^j \leq  \ z_{ub}
\end{alignat}
and 
\begin{alignat}{5}
\label{ctBC1}\sum_{p\in P}\sum_{i\in \{1,\ldots, T_j\}}\tilde{x}_i^p x_i^p \leq N-1
\end{alignat}
to the current MILP.  We then continue the Branch-and-Bound algorithm.
The inequality \eqref{ctBC} updates the upper bound of $z^j$. Adding the inequality \eqref{ctBC1} makes the solution $\tilde{x}$  unfeasible for the MILP. 
The algorithm stops when no solution exists and the best solution is $x^*$ with a noise $\noise(x^*)$. 

Recall that, as mentioned in Section~\ref{sec:noise}, the noise is invariant under circular permutations and, as $\minocc_p>0$ for each $p\in P$, we can always assume that the first pitch is of type $p=1$. Hence, we add the equality 
\begin{alignat}{5}
\label{ctx11} x^1_1 = 1 
\end{alignat}
as a constraint, breaking some of the symmetries an thus reducing the space of feasible solutions. Mind that this can only be done for the computation of the exact noise but not for the approximated noise. 

To break the symmetry, we also propose the following method:
Given the sequence $\tilde{s}$ associated to constraint~\eqref{ctBC1}, we can add $N-1$ constraints~\eqref{ctBC1} associated to each sequence obtained from  $\tilde{s}$ by circular permutation.

\section{Implementation and experimental results}\label{sec:results}
In order to compare experimental results produced by the different methods, the tests have been carried out using the same parameters as for the case of Genetic Algorithms (GA), studied in~\cite{becker2006genetic}:

\begin{itemize}
\item There are three different pitch types of length ratios 1, 1.25 and 1.5, respectively.
\item  The height of a pitch is $h=100$.
\item The groove is $q=0.1$. 
\end{itemize}

Notice that these parameters correspond to realistic values, the same pitch types have been studied in~\cite{chiu2002}, for the Goodyear patent described in~\cite{landers82}. 

To solve the TNOP with GA \cite{becker2006genetic}, JAVA is used for the implementation.
The main parameters of the GA are: a maximal population size of 1500 and a crossover probability of 0.3. The crossover is performed randomly.
The mutation probability is 0.15 and the mutation is also random. Roulette and Ranking selection has been used, with a selection pressure of 0.4.
For performance reasons, the coding of the genes has been realized as final Byte Array instead of a binary encoding.
The value semantics of the genes save memory usage and enable use of the efficient "==" operator.
A chromosome then is composed of genes.

To solve the TNOP with the MILP $(\mathcal{P})$ we used the CPLEX 12.7.1 solver.  The iterative algorithm solving $(\mathcal{P})$ for each tire length $T_j$ was implemented in JAVA. The Branch-and-Cut algorithm was implemented in JAVA using CPLEX and the lazy constraint callback. 
The numerical tests have been carried out on an Intel Core i5 of 3.4 GHz in a Linux environment.

In Tables~\ref{tab:resultats}, ~\ref{tab:resultats2} and  \ref{tab:options},  an instance is considered  as a triple $(N, \minocc _p, \maxocc_p)$. According to the tests made with heuristic methods, we do not take into account the constraints~\eqref{ctn1:10}, \eqref{ctn1:7} and \eqref{ctn1:8}. The computational times are given in \textit{sec}, \textit{min:sec} and \textit{hrs:min:sec}.

Table~\ref{tab:resultats} presents some experimental results for the heuristic method (GA) and the MILP with B$\&$C, the latter computing the exact noise.  We can remark that the sequences found coincide up to circular permutation and hence  the solutions found by the heuristic method are also exact. The computational time remains small for GA but increases in a significant way for the MILP with B$\&$C. Indeed, for 10 pitches, the computation times are very close for GA and B$\&$C. However, for 15 pitches the  computation time for B$\&$C ranges from 10 to 25 minutes which still remains acceptable. From 20 pitches on, it increases considerably. 
Notice that the genetic algorithm cannot ensure to find the optimal solution, \textit{e.g.} the noise found for the instance (60, 10, 40) is 3.899 and the one for the instance (60, 1, 58) is 4.132, whereas the noise of the latter should be smaller since its space of feasible solutions is larger than that of the former. 

Table~\ref{tab:resultats2} shows the results found  for the computation of the approximated noise, based on the same instances. The presented values are the following: noise MILP is the approximated noise, real noise is the exact noise of the sequence found by the MILP and the optimal noise is the noise of the optimal sequence, presented in Table~\ref{tab:resultats}. The gap is computed between the real and the optimal noise, with respect to the latter.  
Notice that, for each of the sequences found, the real noise is an upper bound for the optimal noise. 
We can also remark that, from 15 pitches on, the solution found is not optimal but  the computational time is significantly shorter than that of the exact algorithm (MILP with B$\&$C) and the gap might be considered as reasonable. 
Moreover, although the theoretical gap between the real and the optimal noises is bounded by $42\%$, we notice that for our instances the gap computed is much lower, ranging from 3 to 13.2$\%$.

\begin{table}
\centering
\caption{Experimental results -- exact solutions }

\small{
\begin{tabular}{ |c||c||c|c||c|c| } 
\hline
&noise GA =&pitch sequence&time&pitch sequence &time MILP \\
instance& noise MILP& GA & GA & MILP with B$\&$C & with B$\&$C\\
\hline
\hline
(10,1,8)&9.019&1311323331&5&1323331131&7\\
(10,2,6)&9.247&1231123333&3&1233331231&6\\
(10,2,4)&9.268&2213111333&3&1333221311&3\\
(10,3,4)&9.368&1233321123&3&1231233321&2\\
\hline
(15,1,13)&7.027&311113311133312&6&111333123111133&25:27\\ 
(15,2,11)&7.236&112131223333111&5&131223333111112&27:30\\
(15,4,7)&7.261&111222123333111&5&122212333311111&13:13\\ 
(15,4,6)&7.439&332112331112213&4&112331112213332&10:25\\
\hline
(20,6,8) & 6.444& 11123332312321132112  & 26 & 12321132112111233323& 153:53:39 \\
\hline
(60,1,58)&4.131 & 21231123311232313111 & 02:21:34 & -- & --\\
&& 21112311231111131332 &&& \\
&& 13323323333331212131& & &\\
(60,10,40)&3.899 &12311331212133312113 &03:35:87& -- & -- \\
& & 21131112223333311111 & & & \\
& & 23311231333113331121&& &\\
\hline
\end{tabular}}
\label{tab:resultats}
\end{table}


\begin{table}
\centering
\caption{Experimental results -- approximated  solutions }
\begin{tabular}{ |c||c|c|c|c|c|c| } 
\hline
&& & noise & real  & optimal  & gap \\
instance&pitch sequence& time&  MILP&  noise &  noise &  ($\%$) \\
\hline
\hline
(10,1,8)&1121133231& 4&7.100& 9.540 & 9.019 & 5.5\\
(10,2,6)&1121133231& 3&7.100& 9.540 & 9.247  &  3\\
(10,2,4)&1223331113&1&7.177& 9.638 & 9.268 & 3.8\\
(10,3,4)&1332311212&1&7.397& 9.921 & 9.367 & 5.9\\
\hline
(15,1,13)&223333121111131&1:12&5.613& 7.852 & 7.027 & 10.5\\
(15,2,11)&131223333111112&1:11&5.613& 7.852 & 7.236  & 7\\
(15,4,7)&123333131222111&52&5.949& 8.362 & 7.260 & 13.2 \\
(15,4,6)&123333131222111&37&5.949& 8.362 & 7.439  & 11 \\
\hline 
(20,6,8)&22323333113221221111&24:13 &5.094& 6.931 & 6.444 & 7 \\
\hline
\end{tabular}
\label{tab:resultats2}
\end{table}

In Section~\ref{sec:bc}, we proposed two possible improvements due to the invariance of the noise computing under symmetries.   In Table~\ref{tab:options}, we compare the computation times when taking them into account: symmetry 1 corresponds to constraint~\eqref{ctx11} and symmetry 2 to the $N$ constraints coming from circular permutations.  We remark that none of the two symmetries allows to reduce the computation time.

\begin{table}
\caption{Computational times using  symmetries }
\centering
\begin{tabular}{ |c||c|c|c|c| }
\hline
instance & no symmetry  & symmetry 1  & symmetry 2   \\
\hline
\hline
(10,1,8) & 7 & 8 & 10 \\
(10,2,6) & 6 & 7 & 8 \\
(10,2,4) &  3 & 4 &   5 \\ 
(10,3,4) &  2 & 2  & 3  \\
\hline
(15,1,13) & 25:27 & 4:38:48  &   52:12 \\  
(15,4,6) & 10:25 & 1:52:05  & 20:29   \\
\hline
\end{tabular}\\
\label{tab:options}
\end{table}

\section{Conclusion}
We had started in~\cite{smc2016} with a new approach using a MILP for the Tire Noise Optimization Problem. Indeed, the tire shape has to satisfy several constraints \textit{a priori} allowing for linear programming. However, the objective function depends a very non-linear way on the data. In~\cite{smc2016}, we proposed an approximated tire noise model allowing for its linearization.
The approximated noise is guaranteed to differ from the optimal noise by a ratio of at  most 42$\%$.
In the present paper, we take some more constraints into account and  can provide an exact method to minimize the noise thanks to a Branch$\&$Cut algorithm, where we also consider possible symmetries in the pitch sequences.
We have tested all proposed algorithms and compared their performances.
The MILP approach always finds a solution in reasonable time, up to a ratio of 14$\%$, which is much lower than the theoretical one.
The first results on Branch$\&$Cut algorithm show that our algorithm can provide an exact solution for at most 20 pitches.
As an extension, it would be interesting to enhance the optimization part to be able to compute the noise for realistic instances, e.g. by developing a dedicated algorithm or by adding valid inequalities to strengthen the linear relaxation.
This opens a new research field on the tire noise problem.

\bibliographystyle{plain} 
\bibliography{biblioIJOC}

\end{document}